\documentclass[a4paper]{article}
\usepackage{INTERSPEECH2021}
\usepackage{multirow}
\usepackage{booktabs}
\usepackage{caption}
\usepackage{INTERSPEECH2021}
\title{Multi-Level Transfer Learning from Near-Field to Far-Field Speaker Verification}
\name{Li Zhang$^1$, Qing Wang$^1$, Kong Aik Lee$^2$, Lei Xie$^{1^*}$  \thanks{* Corresponding author.}, Haizhou Li$^{3^{\dagger}} $\thanks{$\dagger$ This work was supported by the Science and Engineering Research Council, Agency of Science, Technology and Research, Singapore, through the National Robotics Program under Grant No. 192 25 00054.}  }
\address{
  $^1$Audio, Speech and Language Processing Group (ASLP@NPU), School of Computer Science, Northwestern Polytechnical University, Xi'an, China  \\
  $^2$Institute for Infocomm Research, A$^{\star}$STAR, Singapore  \\
  $^3$Department of Electrical and Computer Engineering, National University of Singapore, Singapore 
 }
\usepackage[colorlinks,linkcolor=blue]{hyperref}            
\email{lizhang.aslp.npu@gmail.com, lxie@nwpu.edu.cn}
\begin{document}
%\footnote{This work was finished in NUS}
\maketitle 
\begin{abstract}
\vspace{-0.1cm}
%it is a common problem that the performance of 
In far-field speaker verification, the performance of speaker embeddings is susceptible to degradation when there is a mismatch between the conditions of enrollment and test speech. To solve this problem, we propose the feature-level and instance-level transfer learning in the teacher-student framework to learn a domain-invariant embedding space. For the feature-level knowledge transfer, we develop the contrastive loss to transfer knowledge from teacher model to student model, which can not only decrease the intra-class distance, but also enlarge the inter-class distance. Moreover, we propose the instance-level pairwise distance transfer method to force the student model to preserve pairwise instances distance from the well optimized embedding space of the teacher model. On FFSVC 2020 evaluation set, our EER on Full-eval trials is relatively reduced by 13.9\% compared with the fusion system result on Partial-eval trials of Task2. On Task1, compared with the winner's DenseNet result on Partial-eval trials, our minDCF on Full-eval trials is relatively reduced by 6.3\%. On Task3, the EER and minDCF of our proposed method on Full-eval trials are very close to the result of the fusion system on Partial-eval trials. Our results also outperform other competitive domain adaptation methods.
%Task1 and Task3 results on full evaluation trials have a bit gap to the fusion system result on 30\% trials.
\end{abstract}
%With our proposed multi-level transfer learning methods, in FFSVC 2020 development trials, we can achieve 22.8\% relative improvement on EER compared with our baseline result in Task1. For task2 we can achieve 38.6\% relative improvement on EER compared with our baseline result. About Task3, we can achieve 32.8\% relative improvement on EER compared with baseline result. In evaluation trials, our full trials test results are even better than 30\% trials test results in top winner's papers in FFSVC 2020 challenge.
\vspace{-0.2cm}
\section{Introduction}

%is defined to determine whether two sets of voice belong to the same speaker. The process of verification 
Speaker verification~(SV) is to decide to accept or reject test utterances according to the enrollment utterances~\cite{bimbot2004tutorial}. In recent years, most speaker verification methods based on deep learning have achieved superior recognition performance under controlled conditions, i.e. close-talk scenarios with less interference and less mismatch. However, their performances drop significantly when the speech is collected in the wild, such as far-field noisy scenarios or mismatch exists. In far-field scenarios,
%one of the common challenges is the mismatch between enrollment and test data.
it is common for users to enroll their voice via close-talking mobile phones and authenticate in the complex far-field daily home environment. So most SV formulas in smart speakers and various voice-enabled IoT gadgets need to deal with domain mismatch between enrollment and test utterances. 

In recent studies, the solutions to domain adaptation in SV tasks can be divided into three categories. The first method is data augmentation, which can make the SV model `see' more acoustic environment variances and obtain more robust speaker embeddings. In far-field speaker verification challenge 2020~(FFSVC 2020)~\cite{Qin}, many systems~\cite{Qin,Novoselov,zhang2020deep,Zhang} have considered data augmentation as a solution for domain adaptation to improve system performances. The second method is to apply adversarial learning to make the distribution of source and target domain more similar~\cite{Wang,Rohdin,Xia,Meng,Luu,li2020speaker}. The third method is to adopt teacher-student~(T/S) model for knowledge transfer learning~\cite{Torrey,Hinton,Chebotar,Sang}. T/S model was firstly introduced to reduce model size by distilling knowledge from a well trained large teacher model to a small student model~\cite{lu2017knowledge}. Moreover, T/S model can deal with domain mismatches by transferring accurate knowledge from teacher model to student model~\cite{movsner2019improving}, which can make the student model robust in different mismatch scenes~\cite{meng2019conditional}.
Besides knowledge transfer learning with the Kullback-Leibler (KL) divergence in T/S model, minimizing the distance from the corresponding embeddings extracted from teacher model and student model can also decrease the mismatch between the teacher and student~\cite{Lin}. Liang et al.~\cite{liang2018learning} and Chen et al.~\cite{chen2020length} proposed invariant representation learning with cosine-based consistency embedding training. Jung et al.~\cite{Jung} proposed the cosine-based T/S to improve short utterance verification performance with the help of long utterances. 
 
However, all previous T/S methods in speech processing only considered classification accuracy guidance from teacher model and the embedding layer mapping between the teacher and student model. The embedding layer mapping aims to reduce the distance of embeddings from the same classes extracted from teacher and student models but ignores enlarging the distance between different classes which is vital as well. Moreover, the methods mentioned above mainly focus on the mismatches between training and test set but we deal with the mismatch between enrollment and test utterances, which is extremely common in far-field speaker verification.

In this paper, we propose the multi-level transfer learning from near-field to far-field to solve the mismatch between enrollment and test utterances. In the proposed method, we make good use of the domain-invariant knowledge from close-talking data to guide our student model to learn with far-field data.  Inspired by the contrastive loss in self-learning~\cite{he2020momentum,chen2020simple,hjelm2018learning}, we develop the contrastive loss to increase the distance between different classes in T/S model. Besides the feature-level knowledge transfer in embedding layer, we propose an instance-level pairwise distance transfer method to force the student model to preserve pairwise instances distance which is calculated from a well optimized embedding space of the teacher model. Experimental results with the proposed method on FFSVC 2020 evaluation trials illustrate that our methods get significant improvements compared 
with several competitive methods.

\section{Teacher-student Framework}
In T/S framework, the source-domain model~(teacher) trained by close-talking data with cross entropy loss aims to generate the corresponding
posterior probabilities as soft labels, which are used in lieu of the hard labels derived
from target-domain model~(student) training by the parallel far-field data. 
\iffalse
The framework of T/S model is in Figure~\ref{fig:T_S_framework}. 
\begin{figure}[th]
\setlength\abovedisplayskip{0cm}
\setlength\belowdisplayskip{0.1cm}
\setlength{\abovecaptionskip}{0.1cm}
\setlength{\belowcaptionskip}{-0.3cm}
  \centering
  \includegraphics [width=\linewidth]{./figs/overview4}
  \caption{The overview of T/S framework.}
  \label{fig:T_S_framework}
\end{figure}
\noindent In this framework, the neural network with dotted line is teacher model.
The teacher model trained by closed-talking data will be fixed the parameters when it assists student model training. 
\fi
Classical T/S learning is to minimize the Kullback-Leibler~(KL) divergence between the output distributions of the teacher network and the student network~\cite{meng2019conditional} given the parallel data $x_{t,i}$ and $x_{s,j}$. The KL-divergence with temperature = $1$~($T=1$) can be written as:
\begin{equation}
\setlength\abovedisplayskip{0.1cm}
\setlength\belowdisplayskip{0.1cm}
\begin{aligned}
     L_{KL}[p(c|x_{t,j};\theta _{t})||p(c|x_{s,i};\theta _{s})] \\
     =\sum_{i=1}^{N}\sum_{c=1}^{C}p(c|x_{t,j};\theta_{t})log[\frac{p(c|x_{t,j};\theta _{t})}{p(c|x_{s,i};\theta _{s})} ], 
\label{DKL}
\end{aligned}
\end{equation}
where $x_{t,j}$ is the $jth$ close-talking utterance which is from the close-talking dataset $X_t=\{x_{t,1},x_{t,2},x_{t,3},...,x_{t,N}\}$. $x_{s,i}$ is the $ith$ far-field utterance which is from the far-field dataset $X_s=\{x_{s,1},x_{s,2},x_{s,3},...,x_{s,M}\}$. We group $x_{t,i}$ and $x_{s,j}$ into the pair $(x_{t,j},x_{s,i})$ and they are respectively input into teacher and student model as well as $x_{t,i}$ and $x_{s,j}$ have the same speaker label. $c$ is the speaker identity which is from the speaker label set $C=\{c_1,c_2,c_3,...,c_c\}$ and $|C|=C$. These probabilities of $p(c|x_{t,j};\theta _{t})$ and $p(c|x_{s,i};\theta _{s})$ are the classification posterior of teacher model and student model respectively. We obtain the best result when temperature equals 1 the same conclusion reported in~\cite{meng2019conditional,lu2017knowledge,jaitly2012application,tan2018knowledge,watanabe2017student}. Cross entropy loss combined with KL divergence is a common loss of knowledge distillation in model compression~\cite{lu2017knowledge} and transfer learning~\cite{meng2019conditional}. The formula of cross entropy is: 
\begin{equation}
\setlength\abovedisplayskip{0.cm}
\setlength\belowdisplayskip{0.1cm}
\begin{aligned}
    L_{CE}(\theta _{s})=-\frac{1}{N}\sum_{i=1}^{N}\sum_{c=1}^{C}c^{'} log(p(c|x_{s,i};\theta _{s})),
\label{LCE}
\end{aligned}
\end{equation}
However, transfer knowledge by the posterior of classification cannot ensure to optimize out a well embedding space of the student model as that of the teacher model. So we propose a multi-level transfer learning method to improve the transfer learning performance on the mismatch problems.

\section{ Multi-Level Transfer Learning}
%feature level structure metric knowledge transfer in embedding space. This view wedges the purpose of speaker verification task that is smaller intra-class distance and larger inter-class distance.
%s特征维度的知识迁移
\subsection{Feature-level knowledge transfer}
\begin{figure}[th]
\captionsetup{font={footnotesize}}
\setlength{\abovecaptionskip}{0.1cm}
\setlength{\belowcaptionskip}{-0.3cm}
  \centering
  \includegraphics [width=\linewidth]{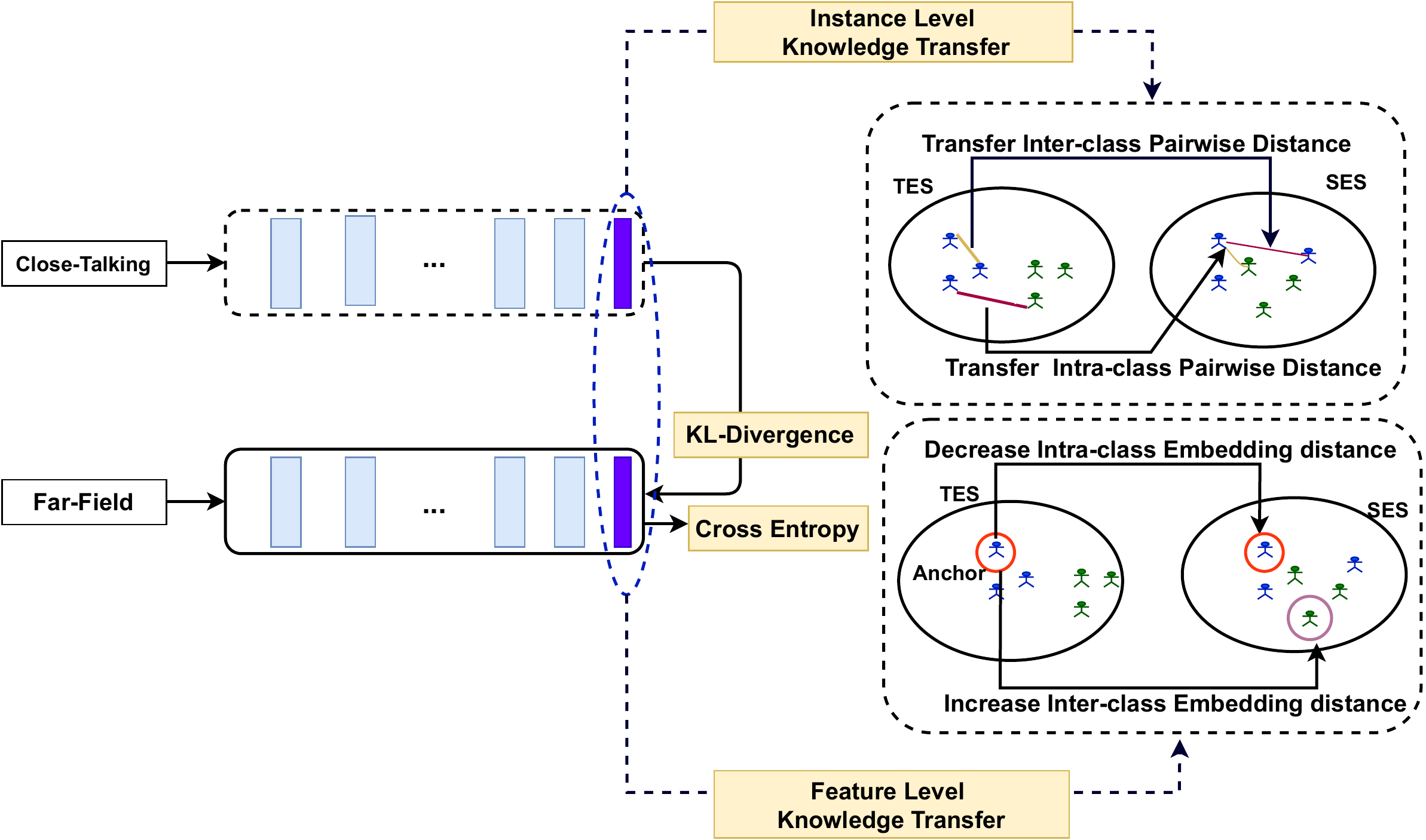}
  \caption{Multi-level transfer learning method in T/S}
  \label{fig:proposed}
\vspace{-0.1cm}
\end{figure}

In order to decrease the mismatch between enrollment and test utterances in far-field speaker verification, we propose a multi-level knowledge transfer method which can not only transfer knowledge in feature-level but also transfer structure information of instance-level pairwise distance. The overview of our proposed method is shown in Figure~\ref{fig:proposed}. The architecture of our method consists of four parts which are teacher model, student model, feature-level transfer learning and instance-level transfer learning. The feature-level and instance-level transfer learning parts operate on the embedding layer. The feature-level transfer learning  aims to increase the inter-class distance as well as to decrease the intra-class distance. The instance-level transfer learning compares the `anchor' embedding extracted from the teacher model with the positive embedding which has the same speaker label as `anchor' embedding and the negative embeddings which have different speaker labels with the `anchor' embedding. And the negative and positive embeddings are extracted from the student model. In the right part of Figure~\ref{fig:proposed}, TES and SES are the abbreviations of teacher embedding space and student embedding space respectively. In the circle, different color speakers mean that they have different speaker labels.
%The multi-level transfer learning method can take good advantage of close-talking data to optimize the student model embedding space of far-field data having the same distribution as teacher model's, which is Unlike previous knowledge transferring by posterior of classification~\cite{Heo} or embedding layer feature mapping~\cite{Jung,Lin}.
 \vspace{-0.2cm}

In previous work, different kinds of metric distance losses are introduced to decrease mismatch discrepancy between the teacher embedding and the student embedding to improve the generalization of student model. The general formulation of different distance losses is:
\begin{equation}
\setlength\abovedisplayskip{0.cm}
\setlength\belowdisplayskip{0.1cm}
\begin{aligned}
    L_F= D(f_{\theta_{t}}(x_{t,i}),f_{\theta_{s}}(x_{s,j})),
\label{feature_level_mapping}
\end{aligned}
\end{equation}
where $f(\cdot)$ is the output embeddings of speaker extractor. $f_{\theta_{t}}(x_{t,i})$ and $f_{\theta_{s}}(x_{s,j})$ are the speaker embeddings extracted from teacher model and student model respectively. $D$ is the general distance measure formulas, such as mean square error~(MSE) loss~\cite{Jung},  maximum mean discrepancy~(MMD) loss~\cite{Lin} or cosine similarity loss~\cite{Jung}. All the losses are calculated on the embedding layer in speaker extractor so we call this kind of methods feature-level knowledge transfer. In this paper, we compare the performance of different distance losses. 

Although the above embedding mapping methods transfer invariant knowledge of the same speaker from teacher model to student model, these methods ignore inter-class distance learning of different speakers. So we propose a new feature-level knowledge transfer learning loss based on the contrastive loss in self-supervise learning. The primary idea of contrastive learning in self-supervised representation learning~\cite{chen2020simple,he2020momentum,hjelm2018learning} is to pull an anchor and a `positive' sample together in embedding space, and push the anchor apart from many `negative' samples similar as the goal of speaker verification.
 
We develop the contrastive loss to transfer knowledge from teacher model to student model. In our task, the anchor of this contrastive loss is the embedding extracted from the well trained teacher model. The positive and negative embeddings are extracted from student model. Since our task is supervised we can calculate the contrastive loss of teacher and student according to the labels of samples. The developed formula is: 
\begin{equation}
\setlength\abovedisplayskip{0.1cm}
\setlength\belowdisplayskip{-0.1cm}
\begin{aligned}
      L_{F^{'}}= -\frac{1}{N}\sum_{i\in N}log\frac{exp(<f_{\theta_{t}}(x_{t,i}), f_{\theta_{s}}(x_{s,j})> )}{ \sum_{a\in A(i)}exp(<f_{\theta_{t}}(x_{t,i}), f_{\theta_{s}}(x_{s,a})>)},
\label{st-contrastive}
\end{aligned}
\end{equation}
where $<,>$ is the inner product of embeddings and $f_{\theta_{t}}(\cdot)$ and $f_{\theta_{s}}(\cdot)$ are embeddings extracted from the teacher model and the student model respectively. $f_{\theta_{t}}(x_{t,i})$ and $f_{\theta_{s}}(x_{s,j})$ have the same speaker label while $f_{\theta_{t}}(x_{t,i})$ and $f_{\theta_{s}}(x_{s,a})$ have different labels. $A(i)$ are the indexes of embeddings extracted from teacher model as well as the speaker labels of $A(i)$ samples are different from $f_{\theta_{t}}(x_{t,i})$'s.  The formula takes the embedding extracted from teacher model as the anchor to optimize the embedding space of student model in two aspects. The first aspect worked by the molecule decreases the distance of embeddings from the same class. The other aspect is to increase the distance of embeddings from different classes. In this way, the trustworthy anchor is extracted from the well trained teacher model.
%A resurgence of work in  This method is feature level knowledge transfer aimed to make student model impersonates teacher model performance. However, the core target of speaker verification task is to optimize embedding space. This inspires us to consider optimize student's  embedding space like the teacher model. So we proposed a innovative sample level structure information transfer in speaker embedding space to make the student model imitate pairwise distance of samples from well trained teacher model. The details of our proposed method is described in following part.
%

%样本维度的知识迁移
\vspace{-0.1cm}
\subsection{Instance-level knowledge transfer}
 
With a well trained teacher model, the embedding space generated by the teacher model has a more reliable reference compared with that of the student.
Ideally, if there is no mismatch between teacher model and student model, they will have the same distribution in the embedding space. Suppose one batch of embeddings extracted from teacher model is $F_t(\theta_{t})=[f_{\theta_t}(x_{t,1}),f_{\theta_t}(x_{t,2}),f_{\theta_t}(x_{t,3}),...,f_{\theta_t}(x_{t,B})]$ with the size $[B, F]$ as the embeddings $F_s(\theta_{s})=[f_{\theta_s}(x_{s,1}),f_{\theta_s}(x_{s,2}),f_{\theta_s}(x_{s,3}),...,f_{\theta_s}(x_{s,B})]$ extracted from student model. Training batch size is $B$ and $F$ is the dimension of embedding. Unlike previous work, we propose to preserve the pairwise instances distance calculated from $F_t(\theta_{t})$ in $F_s(\theta_{s})$. This method can guide student model towards the embedding space of teacher model.  
%In order to keep the student model preserve the pairwise similarity from teacher model, we proposed to  penalty the difference from pairwise similarity matrices of teacher model and student model. In this way, we can force student model preserve the structure information of embedding space from teacher model.  
The formula is:
\vspace{-0.1cm}
\begin{equation}
\setlength\abovedisplayskip{0.1cm}
\setlength\belowdisplayskip{0.cm}
\begin{aligned}
     L_{I}=\frac{1}{B^2}||(d^{s}- d^{t})||_2,
\label{SL1}
\end{aligned}
\end{equation}
where $L_{I}$ is MSE distance of two similarity matrices. The $d^{s}$ and $d^{t}$ are pairwise similarity matrices in one batch of 
embeddings generated by the teacher model and the student model respectively. 
%Suppose $x_i$ and $x_j$ are two instances in one batchsize:
\begin{equation}
\setlength\abovedisplayskip{0.1cm}
\begin{aligned}
      d^{t}= F_t(\theta_{t}) \cdot F_t(\theta_{t})^{T},
\label{SL2}
\end{aligned}
\end{equation}
\begin{equation}
\setlength\abovedisplayskip{0.cm}
\setlength\belowdisplayskip{-0.1cm}
\begin{aligned}
     d^{s}= F_s(\theta_{s}) \cdot F_s(\theta_{s})^{T}.
\label{SL3}
\end{aligned}
\vspace{-0.1cm}
\end{equation}
The operator $\cdot$ denotes matrix multiplication and $T$ is the matrix transposition. The minimization loss in Eq.~(\ref{SL1}) can make student model preserve teacher model's instance-level pairwise distances. This method can penalty the difference from pairwise similarity matrices of the teacher model and the student model so that we can optimize the embedding space of the student model to make it closer to the teacher model's.
%In this way, we can force student model preserve the structure information of embedding space from teacher model.Because this knowledge transfer method is based on instance level so we call our method named instance level knowledge distillation abbreviated  ILKD.
%A graphical illustration which illustrates our method is shown in the upper right corner of Figure \ref{fig:proposed}.
This idea is illustrated in the upper right corner of Figure \ref{fig:proposed}.
% \usepackage{multirow}

\vspace{-0.1cm}
\section{Experiments  }
\vspace{-0.1cm}
\subsection{Datasets }
We verify our method on the FFSVC 2020 dataset~\cite{Qin}. The FFSVC 2020 dataset contains close-talking iPhone recordings and far-field  microphone array recordings. The challenge includes three tasks. Task1 is a text-dependent~(Nihao, Miya) speaker verification with close-talking iPhone recordings for enrollment and far-field microphone array recordings for test. Task2 is a text-independent speaker verification with close-talking iPhone recordings for enrollment and far-field microphone array recordings for test. Task3 is a text-dependent~(Nihao, Miya) speaker verification with close-talking iPhone recordings for enrollment and recordings from far-field distributed microphone arrays recordings for test. Each task exists domain mismatch in development trials and evaluation trials. In FFSVC 2020, the organizer only released the results on 30\% of the evaluation trials, denoted as \textbf{Partial-eval} trials. Since we don't know which 30\% is the partial evaluation trials, we evaluate on all the evaluation trials, denoted as \textbf{Full-eval} trials which should be more challenging. We use openslr\footnote{http://www.openslr.org} datasets~(SLR33, SLR38, SLR47, SLR49, SLR62, SLR82, SLR85) to pretrain the teacher and student models. The datasets for pretraining are augmented with MUSAN~\cite{snyder2015musan} and room impulse response~(RIR)~\cite{alien1976image} datasets by KALDI recipe~\cite{snyder2018x}.
%In FFSVC 2020 challenge, papers released by top winners with 30\% trials test results because official leadboard only shows results of  30\% eveluation trials. In our evaluation results, we test full trials in evaluation because we don't know which 30\% trials was selected to report by official. 

%In our work, we test all trials in evaluation. Our EER on Task2 even better than others' 30\% trials' results. 
\vspace{-0.1cm}
\subsection{Experimental Setup}
We adopt a small size Thin ResNet-34~\cite{heo2020clova} with SE-block~\cite{zhou2019deep} as our speaker verification model. 
%With this structure and key word spotting system we have got the second place~\cite{hou2021npu} in PVTC challenge in 2021~\cite{jia20212020}, a far-field key words spotting and speaker verification union task.  
The channel-wise attention has six times reduction ratio ~\cite{heo2020clova} in the SE-block to decrease the parameters as well as to ensure the model's good performance. We use attention statistic pooling~(ASP)~\cite{heo2020clova} as the pooling layer and adopt the penultimate layer of the network as the embedding extraction layer. Our teacher and student models have the same structure and the details of the model structure is shown in Table~\ref{tab:backbone}. The model has about 8 million parameters and the performance of our baseline system is better than the challenge official baseline, which has 21 million parameters~\cite{Qin}. In our work, we adopt the softmax loss function. The number nodes of the last layer are speaker numbers~(NumSpkrs). Eighty dimensional mel-filter bank features with 25ms window size and 10ms window shift are extracted as model inputs. During training, the batch size of every iteration is 400 with 2 GeForce RTX 3090. The learning rate is initialized from 0.01 and decays by the original 10\% every 2 epochs. The optimizer is stochastic gradient descent~(SGD) in pytorch. 
%Tiny model are easy to deploy and do experiences and worth to advocate.
\begin{table}[th]
\centering
\setlength{\abovecaptionskip}{0.1cm}
\setlength{\belowcaptionskip}{-0.3cm}
\footnotesize
\captionsetup{font={footnotesize}}
\caption{T/S Model Structure }
\label{tab:backbone}
\begin{tabular}{lclclcl}
\hline
Layer    & \multicolumn{2}{c}{Kernel Size}               & \multicolumn{2}{c}{Stride}         & \multicolumn{2}{c}{Output Shape}                       \\ \hline
Conv1    & \multicolumn{2}{c}{$ 3\times 3 \times 32 $}   & \multicolumn{2}{c}{$ 1 \times 1 $} & \multicolumn{2}{c}{$ L \times 80 \times 32 $}          \\ \hline
Res1     & \multicolumn{2}{c}{$ 3 \times 3 \times 32$}   & \multicolumn{2}{c}{$ 1 \times 1 $} & \multicolumn{2}{c}{$ L \times 80 \times 32 $}          \\ \hline
SE-Block & \multicolumn{2}{c}{-}                         & \multicolumn{2}{c}{-}              & \multicolumn{2}{c}{$ L \times 80 \times 32 $}          \\ \hline
Res2     & \multicolumn{2}{c}{$ 3 \times 3 \times 32 $}  & \multicolumn{2}{c}{$ 1 \times 1 $} & \multicolumn{2}{c}{$ L \times 40 \times 64 $}          \\ \hline
SE-Block & \multicolumn{2}{c}{-}                         & \multicolumn{2}{c}{-}              & \multicolumn{2}{c}{$ L \times 40 \times 64 $}          \\ \hline
Res3     & \multicolumn{2}{c}{$ 3 \times 3 \times 64 $}  & \multicolumn{2}{c}{$ 2 \times 2 $} & \multicolumn{2}{c}{$ L_{/2} \times 20 \times 128$} \\ \hline
SE-Block & \multicolumn{2}{c}{-}                         & \multicolumn{2}{c}{-}              & \multicolumn{2}{c}{${ L_{/2} \times 20 \times 128 }$} \\ \hline
Res4     & \multicolumn{2}{c}{$ 3 \times 3 \times 128 $} & \multicolumn{2}{c}{$ 2 \times 2 $} & \multicolumn{2}{c}{${L_{/4} \times 10 \times 256 }$}  \\ \hline
SE-Block & \multicolumn{2}{c}{-}                         & \multicolumn{2}{c}{-}              & \multicolumn{2}{c}{${L_{/4} \times 10 \times 256 }$}  \\ \hline
Flatten  & \multicolumn{2}{c}{-}                         & \multicolumn{2}{c}{-}              & \multicolumn{2}{c}{${L_{/8} \times 2560 }$}           \\ \hline
ASP      & \multicolumn{2}{c}{-}                         & \multicolumn{2}{c}{-}              & \multicolumn{2}{c}{5120}                               \\ \hline
Linear   & \multicolumn{2}{c}{512}                       & \multicolumn{2}{c}{-}              & \multicolumn{2}{c}{512}                                \\ \hline
Softmax   & \multicolumn{2}{c}{-}                       & \multicolumn{2}{c}{-}              & \multicolumn{2}{c}{NumSpkrs}                                \\ \hline
\end{tabular}
\vspace{-0.6cm}
\end{table}

\subsection{Training Strategy}
Firstly, we use the openslr datasets to pretrain a robust out-of-domain speaker verification model. Then we retrain the out-of-domain model with FFSVC 2020 train set to construct our baseline.  After that, we retrain the out-of-domain model with iPhone recordings in FFSVC to get a well trained teacher model. We initialize our student model with the parameters of the baseline model. We group close-talking iPhone recordings and the corresponding recordings in train set of FFSVC 2020 into pairs as the inputs of teacher and student model respectively. Each pair utterances have the same speaker label. The reason we use all of FFSVC 2020 train set instead of only far-field microphones recording data as the inputs of student model is that we want the student model to be robust to close-talking and far-field data at the same time. When we verify each knowledge transfer method, we fix the parameters of well trained teacher model, and just train the student model. In training process, each knowledge transfer loss will be combined with the softmax loss to maintain the discriminative power~\cite{Jung} of student model.
%The balance hyparameter between cross entropy loss and knowledge transfer loss is set according to the proportion of cross entropy and knowledge transfer loss. 
In our task, our proposed method combined with cross entropy loss is formulated as:
\vspace{-0.1cm}
\begin{equation}
\setlength\abovedisplayskip{0.1cm}
\setlength\belowdisplayskip{0.0cm}
\begin{aligned}
    L_{CE\_F^{'}} = L_{CE}+ \lambda_{1} \times L_{F^{'}},
\label{LCE3}  
\end{aligned}
\end{equation}
\begin{equation}
\begin{aligned}
\setlength\abovedisplayskip{-0.1cm}
\setlength\belowdisplayskip{-0.1cm}
    L_{CE\_I} = L_{CE}+ \lambda_{2} \times L_{I},
\label{LCE2}
\end{aligned}
\end{equation}
\begin{equation}
\begin{aligned}
\setlength\abovedisplayskip{0cm}
\setlength\belowdisplayskip{0.cm}
    L_{CE\_F^{'}\_I} = L_{CE}+  \lambda_{1} \times L_{F^{'}}+ \lambda_{2} \times L_{I}.
\label{LCE1}
\end{aligned}
\end{equation}
 We finetune the hyparamters $\lambda_{1}$ and $\lambda_{2}$ with different groups of values, but experiments show the model can optimize well when we set $\lambda_{1}$ and $\lambda_{2}$ to 0.1 and 10. The feature-level knowledge transfer loss is quite larger than the instance-level knowledge transfer loss, because the number of negative samples far exceeds the number of positive samples. However, excessive penalties for negative samples will make the model unfriendly to positive samples, so $\lambda_{1}$ is 
quite smaller than $\lambda_{2}$. 
%In our task, we set $\lambda_{1}$ and $\lambda_{2}$ to 0.1 and 10 empirically.
%All of hyperparameters setting in multi-loss combination follows each loss has the similar proportion so the optimization can work well. The experiments of reproducing previous methods also follow this principle to tuning hyperparameters in the compound loss function.
 \vspace{-0.1cm}
\section{Results and Analysis}
\subsection{Experimental Results}
In order to test the degree of the mismatch between enrollment and test utterances, we re-organize the utterances from FFSVC 2020 development trials of Task1 into other two set trials. One set of trials is close-talking iPhone recordings as the enrollment as well as close-talking iPhone recordings as the test. Another set of trials is far-field microphones recordings as the enrollment as well as far-field microphones recordings as the test. The above two new sets of trials have no mismatch between the enrollment and test utterances and they are named Close-talking trials and Far-field trials respectively.

% fine-tuning with domain data~\cite{Qin}
We compare our methods with several recent T/S transfer learning methods, i.e. KL-divergence~\cite{Heo}, cosine distance~\cite{Jung} and MMD~\cite{Lin} loss. In order to maintain its discriminative power of the student model, each knowledge transfer learning methods in experiments is collaborate with the softmax loss in training. We use the cosine distance for scoring. From Table~\ref{tab:Task1}, on Task1 development trials, we can observe that the proposed feature-level transfer learning method achieves relative 16.8\%/17.3\% reductions in EER/minDCF compared with our baseline results. And the instance-level transfer learning method achieves relative 18.5\%/15.9\% reductions in EER/minDCF compared with our baseline. With the help of both feature-level and instance-level transfer learning, we achieve relative 22.8\%/27.8\% EER/minDCF 
reductions as compared with our baseline. Meanwhile, our method outperforms others' released by the top winners of FFSVC 2020~\cite{Novoselov,zhang2020deep}. 

Experimental results in Table \ref{tab:task2_task3} show that we achieve 38.6\%/26.3\% relative reductions on EER/minDCF compared with our baseline results on Task2 development trials. On Task3, we achieve 32.8\%/30.1\% relative reductions on EER/minDCF compared with our baseline result.  Compared with the released results of challenge top winners' papers~\cite{Novoselov, zhang2020deep}, our results are better than those of the DenseNet~\cite{zhang2020deep} and the STC system~\cite{Novoselov}. 
\begin{table}[th]
\setlength{\abovecaptionskip}{0.1cm}
\setlength{\belowcaptionskip}{-0.3cm}
\footnotesize
\captionsetup{font={footnotesize}}
\caption{Experimental Results on Task1 Development Trials}
\resizebox{\linewidth}{!}{
\label{tab:Task1}
\begin{tabular}{llllllll}
\hline
                                    & \multicolumn{2}{c}{Task1.trials}                      & \multicolumn{2}{c}{Close-talking.trials} & \multicolumn{2}{l}{Far-field.trials}                 \\ \cline{2-7} 
\multirow{-2}{*}{Methods}           & EER~(\%)                                  & minDCF         & EER~(\%)                & minDCF              & \multicolumn{1}{l}{EER~(\%)} & \multicolumn{1}{l}{minDCF} \\ \cline{1-7} 
Official Baseline~\cite{Qin}                   & 6.30                                 & 0.640          & -                  & -                   & -                       & -                          \\
Our Baseline                        & 5.18                                 & 0.669          & 4.04               & 0.391               & 7.51                    & 0.646                      \\
Teacher Model                       & -                                    & -              & 2.55               & 0.305               & -                       & -                          \\
Student Model                       & 5.18                                 & 0.669          & 4.04               & 0.392               & 7.51                    & 0.646                      \\
KL divergence~\cite{Heo}                       & 4.92                                 & 0.551          & 2.77               & 0.315               & 4.96                    & 0.485                      \\
Cosine Distance~\cite{Jung}                     & 4.71                                 & 0.526          & 2.69               & 0.315               & 4.84                    & 0.475                      \\
MMD~\cite{Lin}                                 & 4.69                                 & 0.530          & 2.36               & 0.329               & 4.26                    & 0.449                      \\
DenseNet~\cite{zhang2020deep}                            & 4.57                                 & 0.49           & -                  & -                   & -                       & -                          \\
STC System~\cite{Novoselov}                       & 4.24                                 & 0.490          & -                  & -                   & -                       & -                          \\ \hline
Feature-Level Knowledge Transfer~(F)    & 4.31                                 & 0.553          & 2.37               & 0.329               & 4.26                    & 0.449                      \\
Instance-Level Knowledge Transfer~(I)   & 4.22                                 & 0.562          & 2.22               & 0.326               & 3.95                    & 0.461                      \\
Multi-level Transfer Learning~(F\&I) & {\textbf{4.00}} & \textbf{0.483} & \textbf{2.01}      & \textbf{0.281}      & \textbf{3.89}           & \textbf{0.426}             \\ \hline
\end{tabular}
}
\vspace{-0.3cm}
\end{table}
\begin{table}[th]
\setlength{\abovecaptionskip}{0.1cm}
\setlength{\belowcaptionskip}{-0.3cm}
\footnotesize
\captionsetup{font={footnotesize}}
\caption{Experimental Results on Task2 and Task3 Development Trials}
\resizebox{\linewidth}{!}{
\label{tab:task2_task3}
\begin{tabular}{llllll}
\hline
                                    & \multicolumn{2}{c}{Task2.trials}                      & \multicolumn{2}{c}{Task3.trials} \\ \cline{2-5} 
\multirow{-2}{*}{Methods}           & \multicolumn{1}{c}{EER~(\%)}              & minDCF         & EER~(\%)            & minDCF          \\ \cline{1-5} 
Official Baseline~\cite{Qin}                    & 6.23                                 & 0.650          & 5.82           & 0.710           \\
Our Baseline                        & 6.80                                 & 0.745          & 4.71           & 0.617           \\
KL divergence~\cite{Heo}                         & 6.60                                 & 0.735          & 4.30           & 0.509           \\
Cosine Distance~\cite{Jung}                       & 6.56                                 & 0.709          & 4.14           & 0.505           \\
MMD~\cite{Lin}                                   & 6.55                                 & 0.704          & 4.10           & 0.504           \\
DenseNet~\cite{zhang2020deep}                           & 4.57                                 & 0.490          & 4.12           & 0.450           \\
STC System~\cite{Novoselov}                       & 4.46                                 & \textbf{0.484} & 3.35           & 0.458           \\ \hline
Feature-Level Knowledge Transfer~(F)    & 4.46                                 & 0.614          & 3.59           & 0.497           \\
Instance-Level Knowledge Transfer~(I)   & 4.62                                 & 0.650          & 3.61           & 0.518           \\
Multi-level Transfer Learning~(F\&I) & {\textbf{4.17}} & 0.549          & \textbf{3.16}  & \textbf{0.426}  \\ \hline
\end{tabular}
}
\vspace{-0.3cm}
\end{table}

The experimental results on FFSVC 2020 evaluation trials are shown in Table~\ref{tab:eval}. The Full-eval in the column of trials means all of the evaluation trials and Partial-eval means 30\% the evaluation trials. Compared with the result of Partial-eval trials with DenseNet~\cite{zhang2020deep} on Task1, although our EER on Full-eval trials has 0.14\% gap, our minDCF is relatively reduced by 6.3\%. On Task2, our result of Full-trials is even better than the result of Partial-eval trials with the fusion system~\cite{Novoselov}. On Task3, our result on Full-eval trials is better than the result of DenseNet~\cite{zhang2020deep} on Partial-eval trials. Moreover, on Task3, our result on Full-eval trials is very close to result of the fusion system~\cite{Novoselov} on Partial-eval trials. 
\vspace{-0.1cm}
\subsection{Analysis}
\begin{figure}[th]
\footnotesize
\captionsetup{font={footnotesize}}
\setlength{\abovecaptionskip}{0.1cm}
\setlength{\belowcaptionskip}{-0.3cm}
  \centering
  \includegraphics [width=\linewidth]{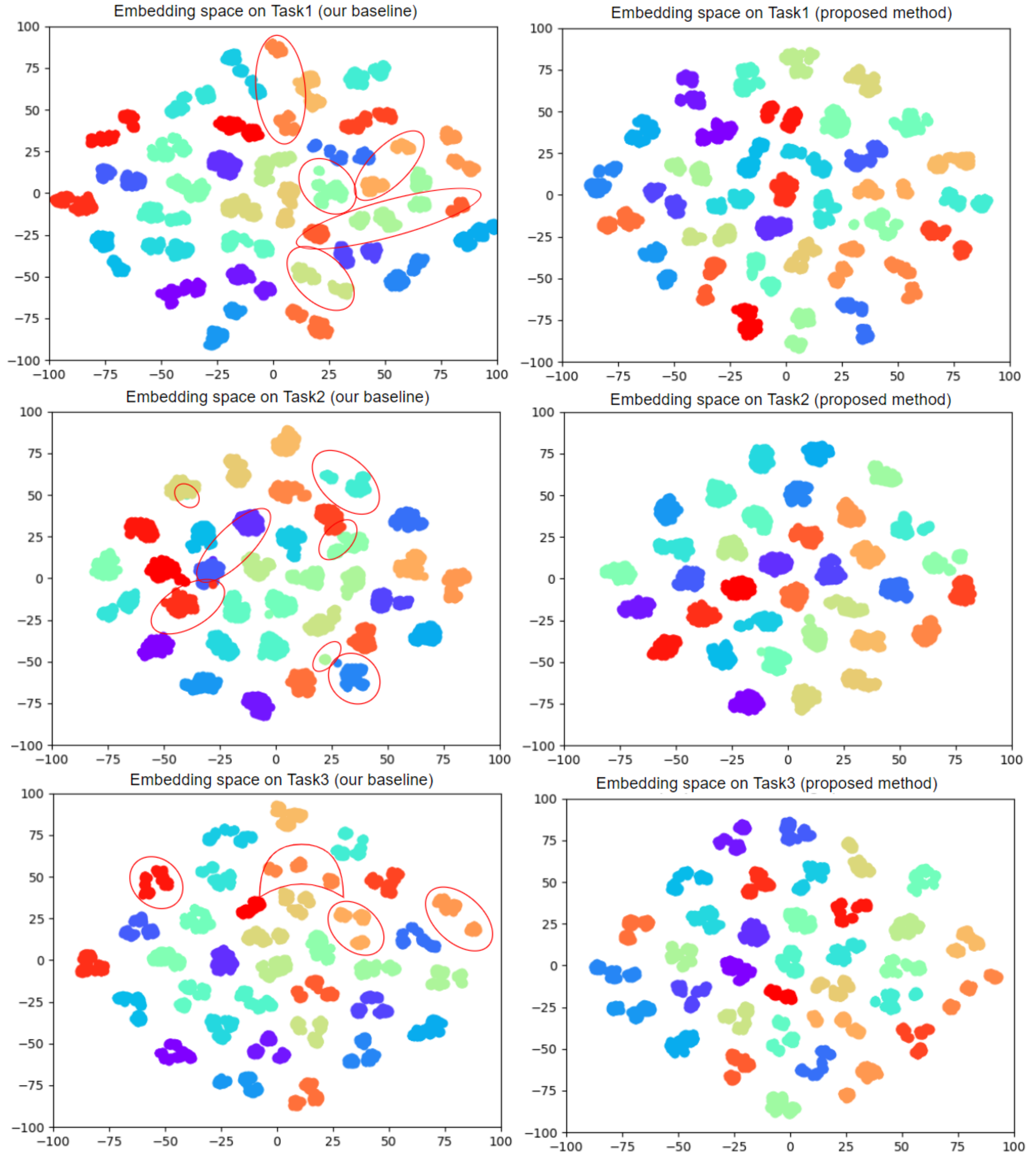}
  \caption{visualization of embedding distribution}
  \label{fig:analysis}
\vspace{-0.2cm}
\end{figure}

We randomly select 35 speakers from evaluation trials for further analysis. We randomly choose 500 embeddings of each speaker to visualize the embedding distribution with TSNE~\cite{van2008visualizing}. The left three pictures in Figure \ref{fig:analysis} illustrate the embedding distribution without our proposed method. We can observe the left visualized embedding space is more messy with red circle line marking the misclassified categories. The right three pictures show the embedding distribution with our proposed multi-level transfer learning method. It is obvious that the right embedding distribution boundary of different classes is more clear and accurate especially on Task2.

% \usepackage{multirow}
\begin{table}[th]
\setlength{\abovecaptionskip}{0.1cm}
\setlength{\belowcaptionskip}{-0.3cm}
\footnotesize
\captionsetup{font={footnotesize}}
\caption{Experimental Results on Evaluation Trials}
\resizebox{\linewidth}{!}{
\begin{tabular}{lccccccc}
\hline
                                    & \multicolumn{2}{c}{Task1.trials}                      & \multicolumn{2}{c}{Task2.trials} & \multicolumn{2}{l}{Task3.trials}                     & \multirow{2}{*}{Trials} \\ \cline{2-7} 
\multirow{-2}{*}{Methods}           &EER~(\%)                                  & minDCF         & EER~(\%)            & minDCF          & \multicolumn{1}{l}{EER~(\%)} & \multicolumn{1}{l}{minDCF} & -                                     \\ \cline{1-8} 
Official Baseline~\cite{Qin}                  & 9.93                                 & 0.870          & 11.87          & 0.950           & 10.68                   & 0.860                      & Partial-eval                                 \\
Our Baseline                 & 6.51                                 & 0.662          &7.85           &0.793         & 6.71                   & 0.610                       & Full-eval                                 \\
DenseNet~\cite{zhang2020deep}                            & 5.78                                 & 0.570          & -              & -               & 6.02                    & 0.530                      & Partial-eval                                   \\
Fusion System~\cite{Novoselov}                       & 5.08                                 & 0.500          & 5.39         &    0.541          & 5.53                    & 0.458                      & Partial-eval                                  \\ \hline

Multi-level Transfer Learning~(F\&I) & {\textbf{5.92}} & \textbf{0.534} & \textbf{4.64}  & \textbf{0.547}  & \textbf{5.64}           & \textbf{0.492}             & Full-eval                                 \\ \hline
\end{tabular}
}
\vspace{-0.5cm}
\label{tab:eval}
\end{table}
\iffalse
our method can achieve the most high improvement. Contrastive transfer learning loss can achieve relative 16.9\% improvement in EER on Task1. And pairwise distance loss can achieve about 18.5\% improvement. After with the two methods combination, we can achieve near 22.8\% relative improvement. Our test trials has no any augmentation, just one utterance enrollment, and one microphone array data test. This verify our methods effect on mismatch problems in SV system. The gap between iPhone.trials and Microphones.trials decreased from 2.2\% to 1.8\% in EER.

At the same time we test our methods on Task2~(Text-independent Microphone Array), Task3~(Text-dependent Distributed Microphones Arrays ) trials. The results showed in Table \ref{tab:task2_task3}. From the table, we can observe that all results have improvements. But for task2, text-independent task, the contrastive transfer learing loss and pairwise loss are more better than MMD \cite{Lin} and Cosine similarity \cite{Jung} loss. We infer that text-dependent task is more easy than text-independent task, it is more easy to mining the information by directly mapping middle layer feature in neural network. For text-independent task, it is more hard to dig out the intra-class similarity, so our method contrastive loss can borrow the 
contrast in one batch size sample thereby to get more information. Our pairwise distance loss can transfer the structure information in teacher embedding space so this can keep the completeness of knowledge. 
\fi
\section{Conclusions}
In this paper, we propose a multi-level transfer learning method to eliminate the mismatch between enrollment and test utterances. In particular, we develop the contrastive loss to reduce the intra-class distance and enlarge the inter-class distance simultaneously. In addition, we propose the instance-level transfer learning to make the student model preserve the instances' pairwise distance in the embedding space of teacher model. With our proposed method, experimental results on FFSVC 2020 development set show that the EERs are lowered relatively by 22.8\%, 38.6\%, 32.8\% on Task1, Task2, and Task3 respectively compared with our baseline results. On FFSVC 2020 evaluation set, our result on Full-eval trials is even better than the fusion system result released by the challenge top winner on Partial-eval trials on Task2. 
\bibliographystyle{IEEEtran}

\bibliography{template}

% \begin{thebibliography}{9}
% \bibitem[1]{Davis80-COP}
%   S.\ B.\ Davis and P.\ Mermelstein,
%   ``Comparison of parametric representation for monosyllabic word recognition in continuously spoken sentences,''
%   \textit{IEEE Transactions on Acoustics, Speech and Signal Processing}, vol.~28, no.~4, pp.~357--366, 1980.
% \bibitem[2]{Rabiner89-ATO}
%   L.\ R.\ Rabiner,
%   ``A tutorial on hidden Markov models and selected applications in speech recognition,''
%   \textit{Proceedings of the IEEE}, vol.~77, no.~2, pp.~257-286, 1989.
% \bibitem[3]{Hastie09-TEO}
%   T.\ Hastie, R.\ Tibshirani, and J.\ Friedman,
%   \textit{The Elements of Statistical Learning -- Data Mining, Inference, and Prediction}.
%   New York: Springer, 2009.
% \bibitem[4]{YourName17-XXX}
%   F.\ Lastname1, F.\ Lastname2, and F.\ Lastname3,
%   ``Title of your INTERSPEECH 2021 publication,''
%   in \textit{Interspeech 2021 -- 20\textsuperscript{th} Annual Conference of the International Speech Communication Association, September 15-19, Graz, Austria, Proceedings, Proceedings}, 2020, pp.~100--104.
% \end{thebibliography}

\end{document}